\documentclass[]{aa}  
\usepackage{xcolor}  
\usepackage{natbib}  
\bibpunct{(}{)}{;}{a}{}{,} 

\usepackage[colorlinks = true,
            linkcolor = blue,
            urlcolor  = blue,
            citecolor = blue,
            anchorcolor = blue]{hyperref} 

\usepackage{graphicx}
\usepackage{txfonts}

\let\ACMmaketitle=\maketitle
\renewcommand{\maketitle}{\begingroup\let\footnote=\thanks \ACMmaketitle\endgroup} 

\usepackage{caption}
\usepackage{subcaption}

\begin{document} 

   \title{Mapping the exo-Neptunian landscape\thanks{Table \ref{tab:weights} is only available at the Centre de Données
astronomiques de Strasbourg (CDS) via anonymous ftp to \url{cdsarc.cds.unistra.fr} (\url{130.79.128.5}).}}

   \subtitle{A ridge between the desert and savanna}

    \author{A.~Castro-Gonz\'{a}lez\inst{ \ref{CAB_villafranca}}
    \and
    V. Bourrier\inst{\ref{obs_geneva}}
    \and
    J.~Lillo-Box\inst{\ref{CAB_villafranca}}
    \and 
    J.-B. Delisle\inst{\ref{obs_geneva}}
    \and \\
    D. J. Armstrong\inst{\ref{warwik}, \ref{warwik_2}}
    \and 
    D.~Barrado\inst{ \ref{CAB_villafranca}}
    \and
    A.~C.~M.~Correia\inst{\ref{cfisuc-coimbra},\ref{obs_paris}}
    }
    
    \institute{Centro de Astrobiolog\'{i}a, CSIC-INTA, Camino Bajo del Castillo s/n, 28692 Villanueva de la Ca\~{n}ada, Madrid, Spain\label{CAB_villafranca} \\\email{acastro@cab.inta-csic.es}
    \and
    Observatoire Astronomique de l’Université de Genève, Chemin Pegasi 51b, CH-1290 Versoix, Switzerland\label{obs_geneva}
    \and
    Centre for Exoplanets and Habitability, University of Warwick, Gibbet Hill Road, Coventry, CV4 7AL, UK\label{warwik}
    \and
    Department of Physics, University of Warwick, Gibbet Hill Road, Coventry, CV4 7AL, UK\label{warwik_2}
    \and
    CFisUC, Departamento de F\'isica, Universidade de Coimbra, 3004-516 Coimbra, Portugal\label{cfisuc-coimbra}
    \and
    IMCCE, UMR8028 CNRS, Observatoire de Paris, PSL Universit\'{e}, 77 Av. Denfert-Rochereau, 75014 Paris, France\label{obs_paris}
    }
    
\date{Received 1 June 2024 / Accepted 4 August 2024}

  \abstract
   {Atmospheric and dynamical processes are thought to play a major role in shaping the distribution of close-in exoplanets. A striking feature of such distribution is the Neptunian desert, a dearth of Neptunes on the shortest-period orbits.}
   {We aimed to define the boundaries of the Neptunian desert and study its transition into the savanna, a moderately populated region at larger orbital distances. Our goal was to acquire new insight into the processes that carved out the Neptunian landscape, and to provide the exoplanet community with a framework for conducting studies on planet formation and evolution.}
   {We built a sample of planets and candidates based on the \textit{Kepler} DR25 catalogue and weighed it according to the transit and detection probabilities. We then used the corrected distribution to study occurrences across the period and period-radius spaces.}
   {We delimited the Neptunian desert as the close-in region of the period-radius space with no planets at a 3$\sigma$ level, and provide the community with simple, ready-to-use approximate boundaries. We identified an overdensity of planets separating the Neptunian desert from the savanna (3.2 days $ \lessapprox P_{\rm orb}$ $\lessapprox$ 5.7 days) that stands out at a 4.7$\sigma$ level above the desert and at a 3.5$\sigma$ level above the savanna, which we propose to call the Neptunian ridge. The period range of the ridge matches that of the well-known hot Jupiter pileup ($\simeq$3-5 days), which suggests that similar evolutionary processes might act on both populations. We find that the occurrence fraction between the pileup and warm Jupiters ($f_{\rm pileup/warm}$ = 5.3 $\pm$ 1.1) is about twice that between the Neptunian ridge and savanna ($f_{\rm ridge/savanna}$ = 2.7 $\pm$ 0.5). This indicates either that the processes that drive or maintain planets in the overdensity are more efficient for Jupiters, or that the processes that drive or maintain planets in the warm region are more efficient for Neptunes.}
   {Our revised landscape supports a previous hypothesis that a fraction of Neptunes were brought to the edge of the desert (i.e. the newly identified ridge) through high-eccentricity tidal migration (HEM) late in their life, surviving the evaporation that eroded Neptunes having arrived earlier in the desert. The ridge thus appears as a true physical feature illustrating the interplay between photoevaporation and HEM, providing further evidence of their role in shaping the distribution of close-in Neptunes.}

   
   \maketitle

%
\section{Introduction}

Understanding how planets form and evolve is one of the main goals of exoplanet research. In this regard, much effort has been put into studying the population of planets with orbits shorter than 30 days. The distribution of these close-in planets shows a striking feature, a dearth of Neptunes at short orbital distances that is commonly referred to as the Neptunian desert \citep[]{2011A&A...528A...2B,2011ApJ...727L..44S,2011ApJ...742...38Y,2013ApJ...763...12B,2016MNRAS.455L..96H,2016NatCo...711201L,2016A&A...589A..75M}. Atmospheric escape is thought to play a major role in shaping the desert \citep[][]{2003ApJ...598L.121L,2003Natur.422..143V,2004A&A...418L...1L,2012MNRAS.425.2931O,2015AREPS..43..459T,2019AREPS..47...67O}, eroding the atmospheres of Neptunes and turning them into smaller planets \citep[e.g.][]{2011A&A...529A.136E,2014ApJ...792....1L}. Dynamical processes are also thought to play an important role \citep{2016A&A...589A..75M,2016ApJ...820L...8M,2017AJ....154..192G,2018Natur.553..477B,2018MNRAS.479.5012O,2022AJ....164..234V}. The core accretion theory for planet formation predicts that giant planets can only form at large orbital distances \citep{1996Icar..124...62P,2006ApJ...648..666R,2015ApJ...811...41L,2019ApJ...878...36L}.  These planets can migrate inwards soon after their formation, before the protoplanetary disk has dissipated, in a process called disk-driven migration \citep{1979ApJ...233..857G,1996Natur.380..606L,2016SSRv..205...77B}.  They can also undergo high-eccentricity tidal migration (HEM) processes \citep[]{2003ApJ...589..605W,2008ApJ...686..621F,2008ApJ...686..580C,2011CeMDA.111..105C,2012ApJ...751..119B}, which can occur at any time in a planet's lifetime. While both mechanisms might be acting, HEM processes are thought to play a major role in populating the close-in Jupiter-size population \citep[see][for a detailed review]{2018ARA&A..56..175D,2021JGRE..12606629F}. However, deciphering the migration history of Neptunian planets is more challenging \citep[e.g.][]{2020A&A...635A..37C}.

In order to disentangle the complex interplay between atmospheric and dynamical processes, different observational constraints are necessary. On the one hand, atmospheric escape has been probed through UV spectroscopy \citep[e.g.][]{2015Natur.522..459E,2017A&A...605L...7L,2018A&A...620A.147B} and helium surveys \citep{2018Sci...362.1388N,2018ApJ...855L..11O,2018Natur.557...68S,2018Sci...362.1384A,2023A&A...676A.130G}. On the other hand, dynamical processes have been explored through the study of orbital elements such as the spin-orbit angle \citep[e.g.][]{2018haex.bookE...2T,2018Natur.553..477B,2021A&A...654A.152B,2023A&A...674A.120A} and eccentricity \citep{2020A&A...635A..37C}, since they are considered to be good tracers of the different migration mechanisms \citep[e.g.][]{2012ApJ...754L..36N,2012ApJ...757...18A,2020A&A...643A..25P,2020AJ....160..179M}. The current constraints provide insight into the possible evolutionary paths of individual systems. However, they still do not allow us to reach firm conclusions on the evolution of close-in planets at a population level. 

Several efforts are underway to obtain a large census of atmospheric escape rates \citep[e.g. the NIGHT spectrograph;][]{2024MNRAS.527.4467F} and system obliquities (e.g. the ATREIDES collaboration; Bourrier et al., in prep) of Neptunian planets. Properly interpreting the results of such large-scale surveys requires accurate characterisation of the distribution of close-in planets, which itself can also provide useful constraints. The Neptunian desert opens up into a moderately populated region at longer periods, a region of the parameter space recently identified as the Neptunian savanna \citep{2023A&A...669A..63B}. Determining where and how the transition between the desert and savanna occurs, as well as the relative planet occurrence between different features, is key to understanding the overall evolution of Neptunian worlds. In this work, using a population-based approach, we aim to map the boundaries of the Neptunian desert and study its transition into the savanna. In Sect.~\ref{sec:sample_selection_biases_mitigation}, we describe the sample selection and bias mitigation. In Sect.~\ref{sec:oc}, we study the distribution of Neptunes across the orbital period space. In Sect.~\ref{sec:boundaries}, we extend our analysis to a wider radius range and explain how we computed new desert boundaries in the 2D period-radius space. In Sect.~\ref{sec:discussion}, we discuss the new Neptunian landscape as a tracer of close-in planets origins, and we conclude in Sect.~\ref{sec:conclusion}.

\section{Sample selection and bias mitigation}
\label{sec:sample_selection_biases_mitigation}

\begin{table}[]
\renewcommand{\arraystretch}{1.3}
\setlength{\tabcolsep}{11pt}
\caption{Transit and detection probabilities of the \textit{Kepler} DR25 catalogue. The complete table is accessible at the Centre de Données astronomiques de Strasbourg.}
\label{tab:weights}
\begin{tabular}{cccc}
\hline \hline
KOI Name        & $P_{\textrm{transit}}^{-1}$ & $P_{\rm detection}^{-1}$ & $w$   \\ \hline
K00001.01    & 7.978060                      & 1.034912              & 8.256587   \\
K00002.01          & 4.133503                        & 1.035139              & 4.278752  \\
K00003.01 & 14.807579                        & 1.036065              & 15.341618  \\
        K00004.01  & 3.892077                    & 1.037361              & 4.037490  \\
K00005.01  & 7.179468                     & 1.038220             & 7.453869\\
...           & ...                         & ...                 & ...     \\ \hline
\end{tabular}
\end{table}

\begin{figure*}
    \centering
    \includegraphics[width=0.48\textwidth]{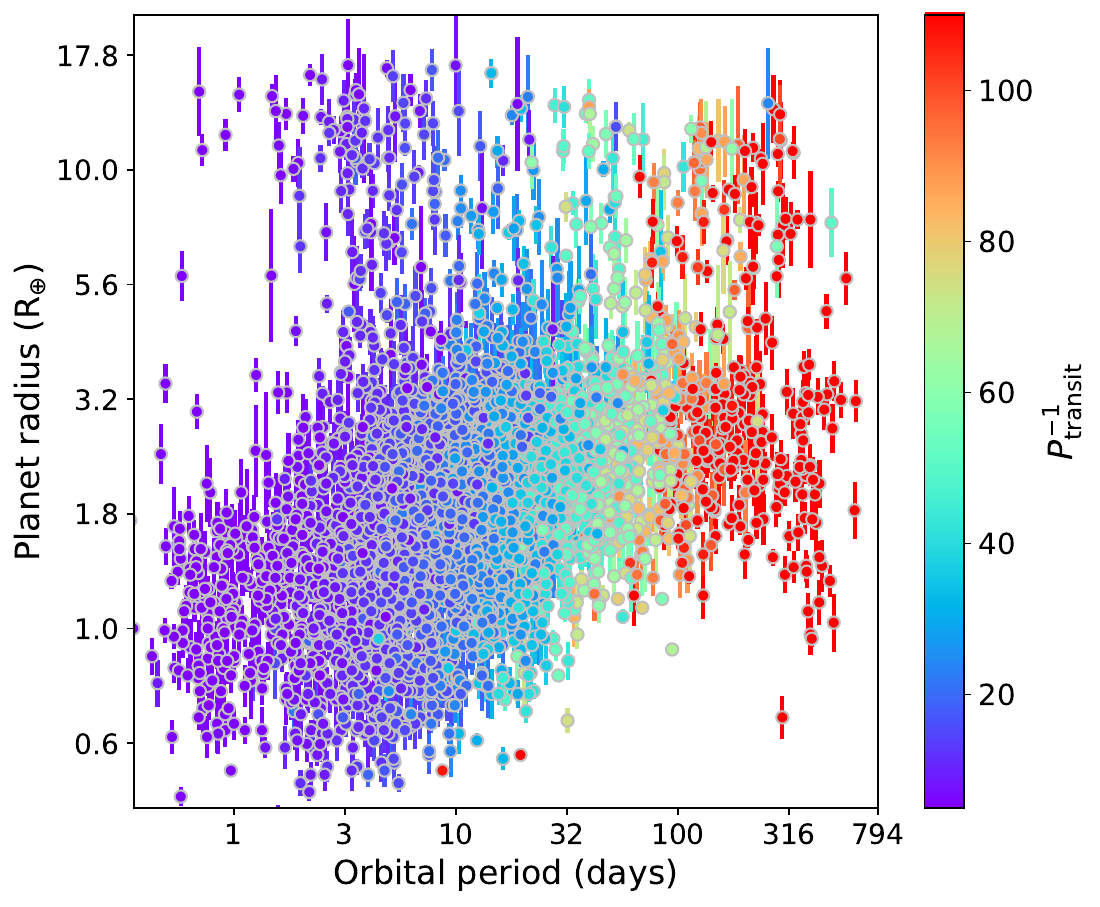}
    \includegraphics[width=0.48\textwidth]{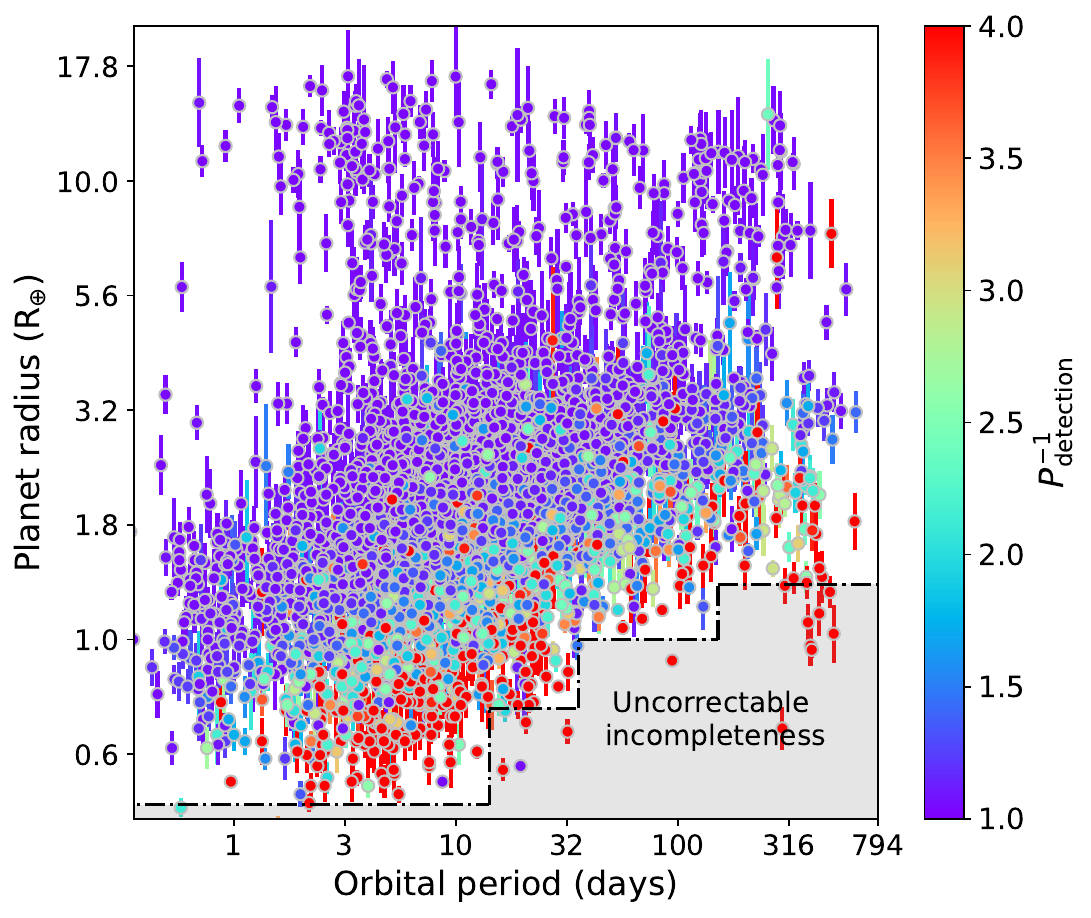}
    \caption{Planet radius as a function of orbital period of the \textit{Kepler} DR25 final catalogue. The colour coding indicates the transit probabilities and the detection probabilities obtained in Sect.~\ref{sec:sample_selection_biases_mitigation}.}
    \label{fig:probs}
\end{figure*}

We built a sample of planets and candidates based on  \textit{Kepler} \citep{2010Sci...327..977B}, since it is the only survey allowing for a complete planet occurrence study ranging from sub-Earths to Jupiters \citep[e.g.][]{2011ApJ...742...38Y}. In particular, we considered the final \textit{Kepler} catalogue \citep[DR25;][]{2018ApJS..235...38T}. This sample is affected by two main observational biases: non-transiting orbital inclinations and insufficient photometric precision. The correction of these biases has been addressed by several works using the inverse detection efficiency method (IDEM), which estimates the probabilities that a particular planet could have been detected orbiting the observed stars \citep[e.g.][]{2012ApJS..201...15H,2012ApJ...753...90M,2013ApJS..204...24B,2013ApJ...766...81F,2013ApJ...767L...8K,2013PNAS..11019273P,2015ApJ...810...95C,2015ApJ...807...45D,2015ApJ...798..112M}. However, these planet-by-planet probabilities are not typically available, and instead occurrence studies provide the planet occurrences in pre-defined period-radius bins. For this work, we followed the IDEM approach to compute such probabilities and made them public to facilitate both the reproducibility of our results and the execution of occurrence studies in other regions of the parameter space. We would like to warn readers, however, that this method has been found to be imprecise near the detection threshold, that is, for small planets at large orbital distances, where there are few detections (see Fig.~\ref{fig:probs}). Hence, we do not recommend using our IDEM probabilities to study occurrences in this region of the parameter space. To that aim, more sophisticated techniques based on extrapolations have been explored \citep[e.g.][]{2014ApJ...795...64F,2014ApJ...791...10M,2018AJ....155..205H,2020AJ....159..248K,2021AJ....161...36B}. 

The geometric probability of observing a transit can be written as a function of the stellar and planetary radii ($R_{\star}$ and $R_{\rm p}$, respectively) and the planetary semi-major axis ($a$):
\begin{equation}
    P_{\textrm{transit}} = \frac{(R_{\star} + R_{\rm p})}{a} \simeq \frac{R_{\star}}{a}.
\end{equation}
In Table~\ref{tab:weights}, we include the inverse of these probabilities $P_{\textrm{transit}}^{-1}$ for the \textit{Kepler} DR25 catalogue. In Figure \ref{fig:probs}, we show the period-radius diagram colour-coded according to $P_{\textrm{transit}}^{-1}$, which indicates the expected dependence with the orbital period. 

The probability of detecting a transiting planet depends on several factors, namely, the instrumental precision, stellar properties (e.g. size, brightness, and activity), and planetary properties (e.g. radius and orbital period). The combination of all of these factors produces a transit signal whose strength is typically quantified by means of its signal-to-noise ratio (S/N). We adopted the definition introduced by the \textit{Kepler} team, since it allows us to take advantage of the photometric noise measured in different timescales by the \textit{Kepler} pipeline \citep{2010ApJ...713L..87J}:
\begin{equation}
    \rm S/N = \frac{\delta}{\sigma_{CDPP}} \sqrt{N},
\end{equation}
where $\delta$ is the transit depth ($\delta$ = $R_{\rm p}^{2} / R_{\star}^{2}$), $N$ is the number of observed transits, and $\sigma_{\rm CDPP}$ is the combined differential photometric precision (CDPP), which indicates the empirical root-mean-square along the transit duration\footnote{The \textit{Kepler} pipeline computes $\sigma_{\rm CDPP}$ in timescales ranging from 1.5 to 15 hours, so we chose the $\sigma_{\rm CDPP}$ corresponding to the timescale closest to the measured transit duration.}. 

The \textit{Kepler} pipeline considers a transit-like signal as a planet candidate if it meets the criterion S/N $>$ 7.1. However, not all planets that meet this criterion are detected with 100$\%$ probability. \citet{2015ApJ...810...95C,2020AJ....160..159C} performed injection-recovery tests and find that the signal recoverability of the \textit{Kepler} pipeline is well described by a $\Gamma$ cumulative distribution, which has the form
\begin{equation}
    p\big(x|a,b,c\big) = \frac{c}{b^{a} \Gamma \big(a\big)} \int_{0}^{x} t^{a-1} e^{-t/b}.
\end{equation}
For this work, we considered the best-fit coefficients $a$ = 33.54, $b$ = 0.2478, and $c$ = 0.9731 as computed by \citet{2020AJ....160..159C} for the DR25 catalogue. This corresponds to a 19.8$\%$ recovery rate of signals with S/N =  7.1, in contrast to the assumed 50$\%$ rate in earlier \textit{Kepler} occurrence works.

For each planet, the fraction of stars around which it would have been detected can be written as
\begin{equation}
    P_{\rm detection} = \frac{\sum_{i=1}^{N_{s}}p_{i}}{N_{s}},
\end{equation}
where $N_{s}$ stands for the number of observed stars. In Table~\ref{tab:weights}, we include the obtained values of $P_{\rm detection}$. In Fig.~\ref{fig:probs}, we plotted the period-radius diagram colour-coded according to $P_{\rm detection}$. 

We corrected each detection for its incompleteness (both geometric and detectability) by assigning it a weight,
\begin{equation}
    w = P_{\rm transit}^{-1} \times P_{\rm detection}^{-1} ,
\end{equation}
which we also include in Table~\ref{tab:weights}. We note that this approach allowed us to obtain a complete sample in the regions of the period-radius space where the detection probability is low yet several detections have been achieved under favourable stellar conditions (i.e. low $\sigma_{\rm CDPP}$). However, the sample remains incomplete in the regions with low detection probabilities and no detections. This is the case for the lower-right region of the period-radius diagram. Building a complete sample involves a trade-off between larger orbital periods and smaller planetary radii. As our aim here is to study the close-in planet population, we selected the \textit{Kepler} detections with $P_{\rm orb}$ < 30 days and $R_{\rm p}$ >~1~$\rm R_{\rm \oplus}$, which allowed us to avoid sparsely populated regions near the detection threshold (see Fig.~\ref{fig:probs}, right panel). We note that a certain fraction of candidates in the \textit{Kepler} DR25 catalogue can be false positives (e.g. brown dwarfs or eclipsing binaries). We explored the potential impact of these false detections using two different approaches. On the one hand, we assigned each planet candidate an additional weight based on its false positive probability (FPP) as estimated by \texttt{vespa}\footnote{The FPPs of the candidates were retrieved from the NASA Exoplanet Archive: \url{https://exoplanetarchive.ipac.caltech.edu/cgi-bin/TblView/nph-tblView?app=ExoTbls&config=koifpp}} \citep{2016ApJ...822...86M}. On the other hand, we restricted our sample to candidates statistically validated as planets. In both cases, the computed occurrences are consistent with those from the original DR25 catalogue considered for this work. This result is in agreement with the work by \citet{2020AJ....159..279B}, who explored the effect of vetting completeness on \textit{Kepler} planet occurrences. The authors found that correcting for reliability can impact the occurrence rates near the detection limit, at orbital periods longer than 200 days and radii smaller than 1.5 $\rm R_{\oplus}$, a parameter space far beyond the population analysed in the present work. 

\section{Occurrence across the orbital period space}
\label{sec:oc}

Transiting planets are commonly classified into three groups according to their radii \citep[e.g.][]{2012ApJS..201...15H,2019PNAS..116.9723Z}: small planets (also known as sub-Neptune planets; $R_{\rm p}$ < 4$\rm R_{\oplus}$), gas giants (also known as Jupiter-size planets; $R_{\rm p}$ > 10$\rm R_{\oplus}$), and intermediate planets (also known as Neptunian planets; 4$\rm R_{\oplus}$ $<R_{\rm p}$ $<$ 10$\rm R_{\oplus}$). In this section, we study planet occurrences across the orbital period space, with the goal of finding where and how the Neptunian desert transitions into the savanna. We aim to study specific features of intermediate planets, so we focus our analysis on a reduced region of the commonly adopted radius range to minimise a potential contamination from the adjacent populations. In Sect.~\ref{sec:oc_nep}, we study the distribution of Neptunian planets with radii 5.5$\rm R_{\oplus}$ $<R_{\rm p}$ $<$ 8.5$\rm R_{\oplus}$. For completeness, in Sect.~\ref{sec:oc_giant_small} we examine the Jupiter-size ($R_{\rm p}$ > 10$\rm R_{\oplus}$) and sub-Neptune planet regimes ($R_{\rm p}$ < 4$\rm R_{\oplus}$), and in Sect.~\ref{oc:trans} we explore the Neptunian regions adjacent to the giant and small planets' regimes, which we refer to as the frontier Neptunian regimes (4$\rm R_{\oplus}$ $<R_{\rm p}$ $<$ 5.5$\rm R_{\oplus}$ and 8.5$\rm R_{\oplus}$ $<R_{\rm p}$ $<$ 10$\rm R_{\oplus}$).

\begin{figure*}
    \centering
    \includegraphics[width=0.4743\textwidth]{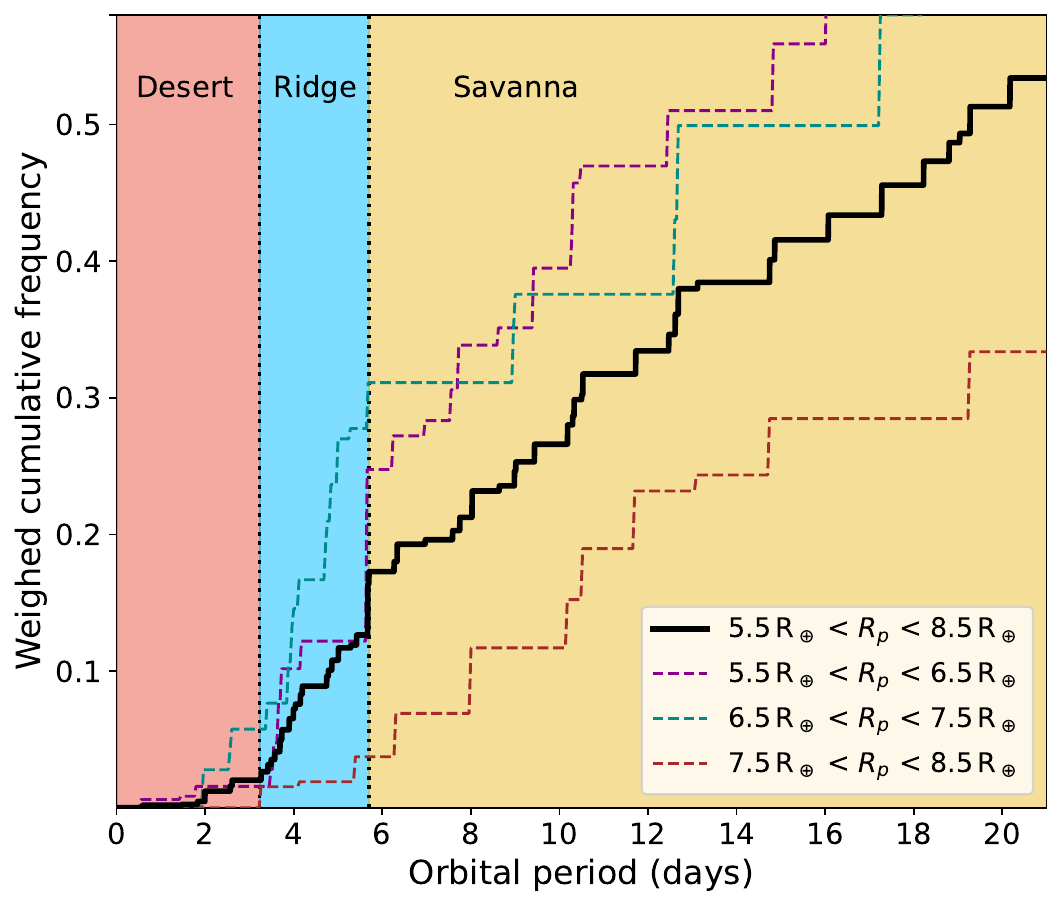}
    \includegraphics[width=0.486\textwidth]{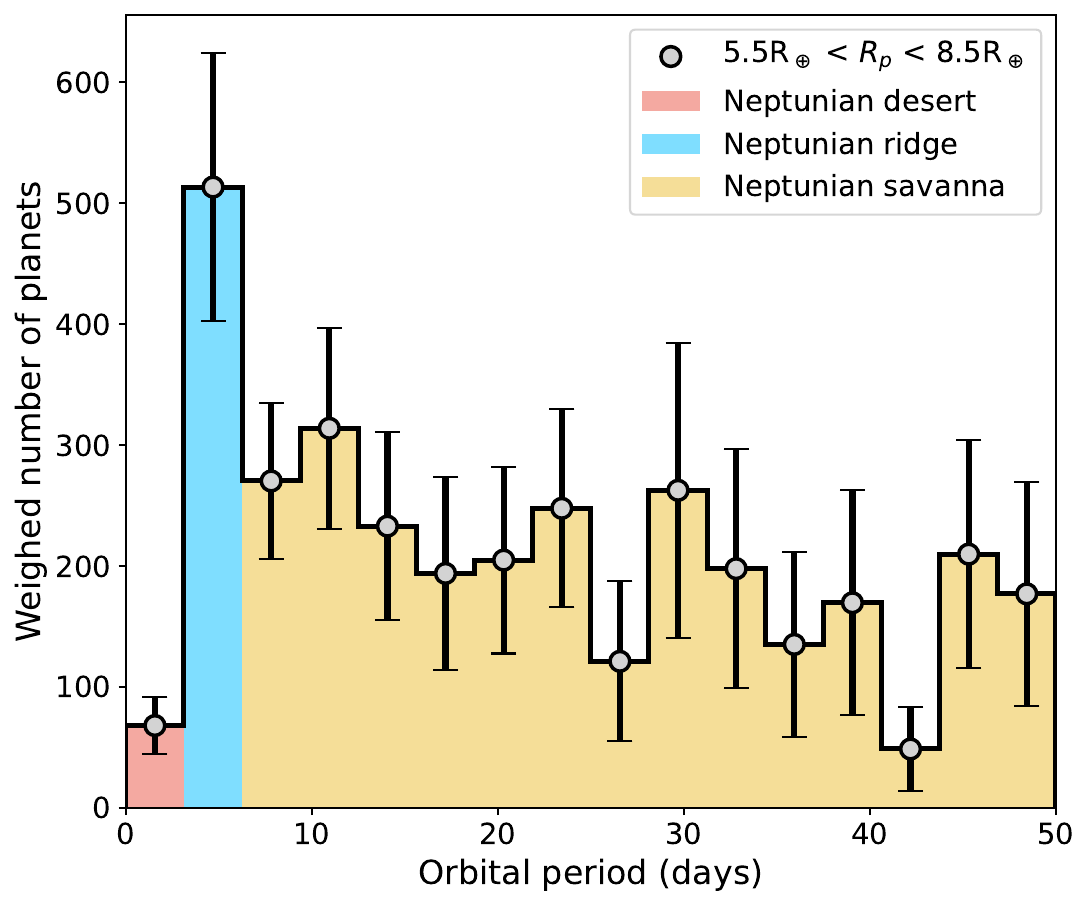}
    \caption{Distribution of Neptunian planets across the orbital period space, where three regimes are differentiated: a significant deficit of planets at periods $\lessapprox$ 3.2 days (i.e the Neptunian desert), a moderately populated region at periods $\gtrapprox$ 5.7 days (i.e. the Neptunian savanna), and an overdensity of planets between these regimes (i.e. the Neptunian ridge). The histogram error bars were computed as the square root of the quadratic sum of the weights. }
    \label{fig:oc:nep}
\end{figure*}

\subsection{Distribution of Neptunian planets}
\label{sec:oc_nep}

In Fig.~\ref{fig:oc:nep}, left panel, we plotted the weighed cumulative frequencies across the orbital period space of Neptunian planets with radii 5.5$\rm R_{\oplus}$ $<R_{\rm p}$ $<$ 8.5$\rm R_{\oplus}$. We find three well-differentiated regimes: a quasi-flat region at $P_{\rm orb} \lessapprox 3.2$ days, a steep frequency increase at 3.2 days $ \lessapprox P_{\rm orb}$ $\lessapprox$ 5.7 days, and a milder increase at $P_{\rm orb} \gtrapprox$ 5.7 days. This distribution validates the existence of the desert and the savanna as two differentiated features, and reveals an abrupt transition between both regimes. This transition corresponds to an overdensity of planets (with respect to both the desert and the savanna), which we propose to name the Neptunian `ridge'. Based on the cumulative frequencies, we built the weighed histogram by considering bin sizes that allow for differentiation between the three regimes (Fig.~\ref{fig:oc:nep}, right panel). This distribution contrasts with the original desert boundaries of \citet{2016A&A...589A..75M}, which range from $\simeq$6 days to $\simeq$15 days (depending on the planet radius) in the Neptunian domain. We note that \citet{2016A&A...589A..75M} mentioned that even though their boundaries intersected at this large orbital period, it was not clear that the desert extended up to it
since the picture was not clear by that time. We also split the Neptunian occurrence into three adjacent radius chunks of 1$\rm R_{\oplus}$ width. We find that the individual distributions of the two chunks in the 5.5$\rm R_{\oplus}$$<$ $R_{\rm p}$ $<$ 7.5$\rm R_{\oplus}$ range peak at the ridge. However, the distribution of the chunk in the 7.5$\rm R_{\oplus}$$<$ $R_{\rm p}$ $<$ 8.5$\rm R_{\oplus}$ range is more homogeneous and does not peak at the ridge. The lower range dominates the complete distribution, and the upper range does not show relevant features. We warn that the number of planets in these individual chunks is too small to reach statistically significant conclusions, but we highlight a possible fading of the ridge in the upper end of the considered radius range. 

We estimated the significance of the Neptunian desert, ridge, and savanna by means of the t-statistic. We propagated the bin uncertainties as $N_{\rm bins}^{-1} (\sum_{i = 1}^{N_{\rm bins}} \delta_{i}^{2})^{1/2}$, with $N_{\rm bins}$ being the number of bins, and $\delta_{i}$ = ($\sum_{j = 1}^{N_{\rm p}} w_{j}^{2})^{1/2}$, where $N_{\rm p}$ is the number of planets detected in each bin. We find that the Neptunian ridge stands out at a 4.7$\sigma$ level above the desert and at a 3.5$\sigma$ level above the savanna. The savanna stands out above the desert at a 4.7$\sigma$ level. We also computed the occurrence fraction between the regimes and find $f_{\rm ridge/desert}$ = 8 $\pm$ 3, $f_{\rm ridge/savanna}$ = 2.7 $\pm$ 0.5, and $f_{\rm savanna/desert}$ = 3.0 $\pm$ 0.9. Overall, both the occurrence fractions and the t-values indicate that the Neptunian ridge describes a region where close-in Neptunes are preferentially found, and it marks the boundary of the Neptunian desert through an abrupt occurrence drop.
\subsection{Distribution of Jupiter-size and sub-Neptune planets}
\label{sec:oc_giant_small}

The distribution of Jupiter-size planets ($R_{\rm p}$ $>$ 10$\rm R_{\oplus}$) has substantial similarities with that of Neptunes. The weighed cumulative frequencies show a steep increase at 3.2 days $\lessapprox$ $P_{\rm orb}$ $\lessapprox$ 5.8 days, which leads to a mild increase at larger orbital periods (Fig.~\ref{fig:oc_jup}). This overdensity in the Jupiter-size domain was noticed shortly after the first discoveries and is commonly known as the hot Jupiter pileup \citep[e.g][]{2007ARA&A..45..397U,2009ApJ...693.1084W}. While similar, the Jupiter-size and Neptunian planet occurrences do show some differences. The occurrence fraction between the hot Jupiter pileup and warm Jupiters at longer orbital periods is larger than for Neptunian planets: $f_{\rm pileup/warm}$ = 5.3 $\pm$ 1.1 versus $f_{\rm ridge/savanna}$ = 2.7 $\pm$ 0.5. Taking the pileup (ridge) density as a reference for warm Jupiters (Neptunes), we find an occurrence ratio between the Neptunian savanna and the warm Jupiter regime of $f_{\rm savanna/warm}$ = 2.0 $\pm$ 0.6.  Another difference is that the hot Jupiter pileup does not abruptly drop to a desert of planets. The distribution of sub-Neptune planets ($R_{\rm p}$ $<$ 4$\rm R_{\oplus}$) is much more homogeneous than that of Neptunian and Jupiter-size planets, and it does not show an overdensity in the 3-5 day period range (Fig.~\ref{fig:oc_sub}).

\subsection{Distribution of frontier Neptunian planets}
\label{oc:trans}

In Fig.~\ref{fig:oc_transitional}, we show the occurrence of frontier Neptunian planets located near the giant and small
planet regimes. On the one hand, the distribution in the 4$\rm R_{\oplus}$ $<R_{\rm p}$ $<$ 5.5$\rm R_{\oplus}$ radius range does not show remarkable features, similarly to the distribution of sub-Neptunes in the upper radius end (see Figs.~\ref{fig:oc_sub} and \ref{fig:oc_transitional}). We note that the number of planets detected in this radius range is large enough (i.e. the bin uncertainties are small enough) to have significantly detected a ridge with the occurrence contrast observed in Sect.~\ref{sec:oc_nep}.  On the other hand, the distribution in the 8.5$\rm R_{\oplus}$ $<R_{\rm p}$ $<$ 10$\rm R_{\oplus}$ radius range shows a clear overdensity of planets at $\simeq$3-5 days, which drops at larger orbital distances (Fig.~\ref{fig:oc_transitional}), similarly to the adjacent Neptunian ridge (see Sect.~\ref{sec:oc_nep}) and hot Jupiter pileup (see Sect.~\ref{sec:oc_giant_small}).

\section{Desert boundaries}
\label{sec:boundaries}

\begin{figure}
    \centering
    \includegraphics[width=0.48\textwidth]{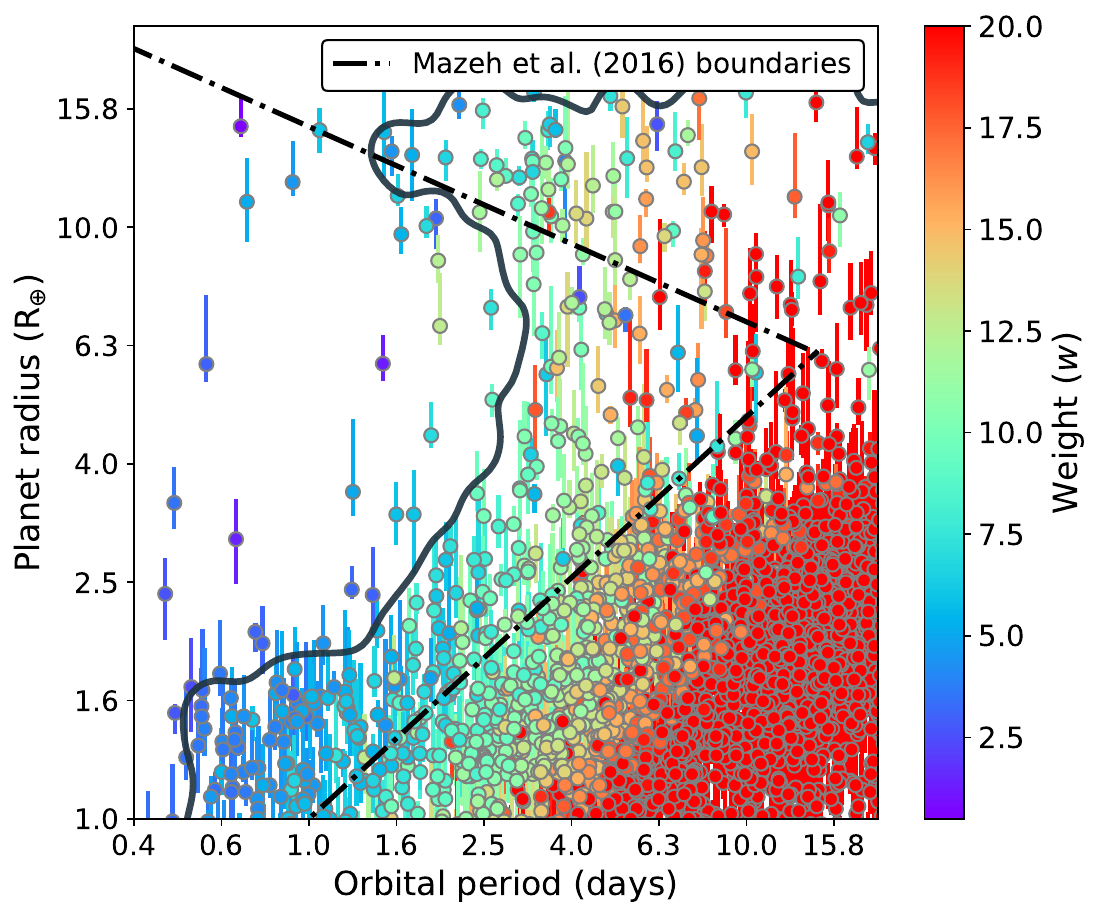}
    \caption{Planet radius versus orbital period of the \textit{Kepler} DR25 catalogue. Each detection is coloured according to the assigned weight to correct for observational biases. The contour line represents the lowest percentile our dataset is sensitive to. }
    \label{fig:our_desert}
\end{figure}

In Sect.~\ref{sec:oc}, we found that the occurrence of Neptunian planets drops to a desert at $P_{\rm orb} \simeq$ 3.2 days. However, the identification of such a desert in the Jupiter-size and sub-Neptune planet regimes cannot be limited to studying occurrences across the orbital period space. In these regimes, the planet distribution at short orbital distances varies strongly as a function of planet radius, so setting a fixed period boundary is not possible.

We aimed to find a population-based desert that is self-consistent in the entire radius range of our \textit{Kepler} sample. To do so, we estimated the probability density function (PDF) of the 2D period-radius distribution corrected for observational biases. We obtained the PDF through a kernel density estimation \citep[KDE;][]{Parzen_1962}. This method is one of the most widely used non-parametric approaches to estimating the underlying PDF of a dataset, and previously has been used to infer planet occurrence distributions \citep[e.g.][]{2014ApJ...791...10M,2023AJ....166..122D}. The KDE method relies upon a functional form (i.e. kernel) that is associated with each data point, and a bandwidth that controls the PDF smoothness. In most cases, the kernel choice has negligible influence in the computed PDF \citep[e.g.][]{2017arXiv170403924C}. We confirmed that this is true for our dataset by testing several kernel options (e.g. Gaussian, uniform, linear, cosine, and exponential), and arbitrarily selected a Gaussian kernel for our final estimation. We considered the rule of thumb introduced by \citet{1986desd.book.....S} to obtain an optimum bandwidth for our dataset. We also tested varying the optimised bandwidth by different factors between 0.5 and 2, and find compatible PDFs. In Fig.~\ref{fig:our_desert}, we plotted the contour line corresponding to the lowest percentile we are sensitive to. For lower percentiles, we find some close-in planets that, depending on their weights, surpass the percentile threshold. Therefore, we consider this contour as a good representation of the desert boundaries. 

We provide the community with the following simple, ready-to-use approximations for the desert boundaries:
\begin{equation}
    \label{eq_6}
    \mathcal{L_{R}} = -0.43 \times \mathcal{L_{P}} + 1.14, \,\, \rm{if} \,\, \mathcal{L_{P}} \in [0.12, 0.47] 
\end{equation}
for the upper limit,
\begin{equation}
    \mathcal{L_{R}} = +0.55 \times \mathcal{L_{P}} + 0.36, \,\, \rm{if} \,\, \mathcal{L_{P}} \in [-0.30, 0.47]
\end{equation}
for the lower limit, and
\begin{equation}
    \mathcal{L_{P}} =+0.47, \,\, \rm{if} \,\ \mathcal{L_{R}} \in [0.61, 0.92] 
\end{equation}
\noindent for the Neptunian domain, where $\mathcal{L_{R}}$ = $\rm log_{10}(R_{\rm p}/R_{\oplus})$ and $\mathcal{L_{P}}$ = $\rm log_{10}(P_{\rm orb}/d)$. The upper and lower boundaries also lead to constant-period limits that we approximate as
\begin{equation}
\label{eq_9}
\mathcal{L_{P}} =\begin{cases} 
 +0.12 & \rm{if} \,\, \mathcal{L_{R}} \in [1.08, 1.20] \\  
 -0.30 & \rm{if} \,\, \mathcal{L_{R}} \in [-0.07, 0.20].  
 \end{cases}
\end{equation}

These boundaries show considerable differences with those derived by \citet{2016A&A...589A..75M}. Our desert is drier in all radius ranges, especially in the sub-Neptune and Neptune domain. Accounting for completeness, we find that 2.2$\%$ of the close-in planet population lies within the \citet{2016A&A...589A..75M} desert, while only 0.1$\%$ lies inside our boundaries. Hence, our desert represents the close-in region of the period-radius space where there are no planets at a 3$\sigma$ level. In addition to the desert dryness, there is also a notable difference regarding the desert shape. Similar to MA+16, our boundaries narrow towards a smaller radius window as the period increases. However, our revised boundaries do not penetrate into the savanna, as the desert is physically limited by the Neptunian ridge. This sets a constant-period boundary in the Neptunian regime at $\simeq$3 days, as we also discuss in Sect.~\ref{sec:oc_nep}.


\section{The Neptunian landscape as a tracer of close-in planets origins}
\label{sec:discussion}

\begin{figure}
    \centering
    \includegraphics[width=0.47\textwidth]{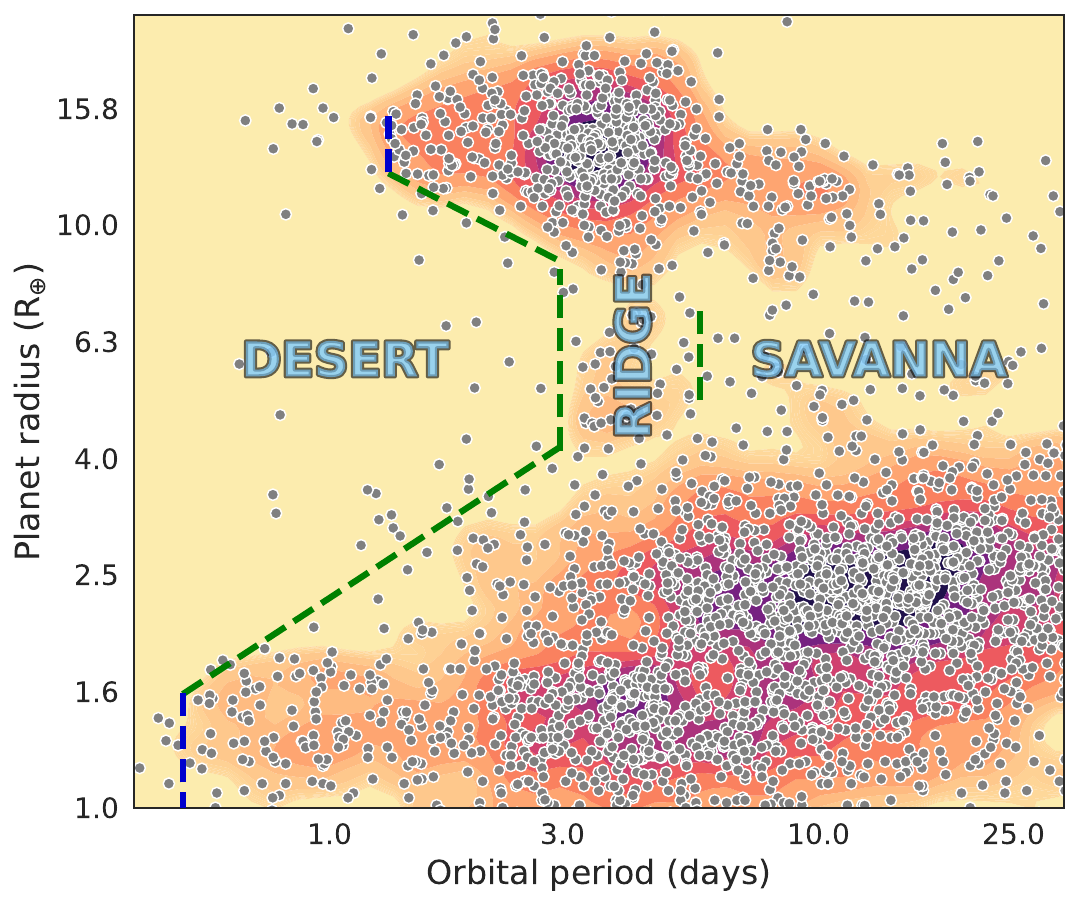}
    \caption{Planet radius as a function of orbital period for all known exoplanets, where we highlight the location of the Neptunian desert, ridge, and savanna derived in this work (Eqs. (\ref{eq_6}) to (\ref{eq_9})). The colour code represents the observed density of planets. This plot has been generated with \texttt{nep-des} (\url{https://github.com/castro-gzlz/nep-des}).}

    \label{fig:Per_Rad}
\end{figure}

In Fig.~\ref{fig:Per_Rad}, we highlight the location of the Neptunian desert, ridge, and savanna (Eqs. (\ref{eq_6}) to (\ref{eq_9})) in the context of all known planets. The ridge spans a period range that coincides with that of the hot Jupiter pileup ($\simeq$3-5 days), which  suggests that similar evolutionary processes might act on both populations. The current observational constraints suggest that both disk-driven migration and HEM processes are needed to explain the observed properties of hot Jupiters. However, a large number of planets in the pileup have been found in elliptical orbits, and the elliptical/circular fraction increases with orbital period, which is interpreted as HEM processes being the main channel populating the pileup \citep[e.g.][]{2018ARA&A..56..175D,2021JGRE..12606629F}. In a recent study, \citet{2020A&A...635A..37C} found that Neptunian planets at the edge of the desert, with $P_{\rm orb}$ $\lessapprox$ 5 days (corresponding to the newly identified ridge), have moderate ($e$ $\lessapprox$ 0.3) but non-zero eccentricities. This suggests that HEM is also the main channel bringing Neptunes to the ridge, and that this migration process might be the main agent populating the $\simeq$3-5 day overdensity observed both in the Jupiter-size and Neptunian populations. While Jupiters in the pileup would remain immune to photoevaporation and eventually have their orbit circularised, the Neptunes observed in the ridge today may have arrived recently enough through HEM that their orbit has not yet circularised and they have survived evaporation \citep{2018Natur.553..477B,2020A&A...635A..37C,2021A&A...647A..40A}. This picture is consistent with additional dynamical and atmospheric constraints, as many Jupiter-size \citep{2012ApJ...757...18A} and Neptune-size \citep{2023A&A...669A..63B} planets in the overdensity have been found on highly misaligned orbits (which is considered a tracer of HEM processes, \citealp[e.g.][]{2012ApJ...754L..36N,2017AJ....154..106N}), and several Neptunian planets within the ridge undergo strong atmospheric escape \citep[e.g.][]{2015Natur.522..459E,2018A&A...620A.147B}. 

In this work, we have found that the occurrence fraction between the hot Jupiter pileup and warm Jupiters is about twice ($f_{\rm pileup/warm}$ = 5.3 $\pm$ 1.1)  that between the Neptunian ridge and savanna ($f_{\rm ridge/savanna}$ = 2.7 $\pm$ 0.5). If we assume that migration processes distribute Jupiter-size and Neptunian planets equally, this result could be explained by photoevaporation, which would be removing Neptunes from the ridge but not from the savanna. However, the reality is probably more complex. The eccentricities of warm Jupiters are preferentially elliptical, while the eccentricities of warm Neptunes in the savanna are preferentially circular\footnote{We refer the reader to Figure 1 of \citet{2020A&A...635A..37C} for an eccentricity comparison between Jupiter-size and Neptunian planets.}. Since photoevaporation does not affect warm Jupiters and is likely inefficient on warm Neptunes in most of the savanna, this implies that HEM processes act differently on warm Jupiter and Neptunes, and/or that HEM and disk-driven migration bring different fractions of Jupiter- and Neptune-size planets from beyond the ice line into the warm planet regime. The warm-Jupiter regime would be preferentially populated through HEM processes, while the Neptunian savanna would be preferentially populated through disk-driven migration. Therefore, the different occurrence fractions between the overdensities and the warm regions of Jupiter-size and Neptunian planets cannot be interpreted through a single process, since the observational constraints hint at a different evaporation and migration mechanisms affecting both populations.

\section{Conclusions}
\label{sec:conclusion}

In this work, we identified an overdensity of intermediate planets (5.5$\rm R_{\oplus}$ $<R_{\rm p}$ $<$ 8.5$\rm R_{\oplus}$) at  3.2 days $ \lessapprox P_{\rm orb}$ $\lessapprox$ 5.7 days, which we call the Neptunian ridge, as it separates the desert and savanna of close-in Neptunes. We also determined accurate, population-based boundaries for the Neptunian desert in the radius-period plane, and here provide the community with simple, ready-to-use approximations for these boundaries. 

The period range of the ridge matches that of the well-known hot Jupiter pileup ($\simeq$3-5 days), suggesting the existence of similar evolutionary pathways populating both regimes. The large number of Jupiter- and Neptune-size planets with eccentric, misaligned orbits in this overdensity further suggests that it is populated primarily by HEM processes. In contrast, the larger fraction of warm Jupiters with eccentric orbits, compared to warm Neptunes on circular orbits in the savanna, suggests that HEM and disk-driven migration act differently on both populations. The different relative fraction of Jupiters and Neptunes in the overdensity, compared to the warm regime, further suggests that HEM is more efficient at bringing Jupiter-size planets closer in, and/or that we only detect Neptunes in the ridge that have survived evaporation because they migrated recently.

These hypotheses must be further tested through large-scale atmospheric and dynamical surveys. Spin-orbit angle surveys such as ATREIDES (Bourrier et al., in prep) will offer further insight into the relative roles of the different migration processes on the Neptunian population, while atmospheric escape surveys such as NIGHT \citep{2024MNRAS.527.4467F} will allow us to determine how deep into the savanna photoevaporation remains efficient, or if the ridge also marks the threshold for the onset of hydrodynamical escape. The results of such surveys coupled with the newly mapped landscape and numerical syntheses of the Neptunian population will provide a clearer picture of the origins and evolution of close-in giants as a whole.

\begin{acknowledgements}
We sincerely thank the referee for the effort and time dedicated to reviewing this manuscript. We are very grateful for the thorough and constructive revisions, which had a positive impact on the results presented in this work. 
A.C.-G. is funded by the Spanish Ministry of Science through MCIN/AEI/10.13039/501100011033 grant PID2019-107061GB-C61. 
This work has been carried out in the frame of the National Centre for Competence in Research PlanetS supported by the Swiss National Science Foundation (SNSF). This project has received funding from the European Research Council (ERC) under the European Union's Horizon 2020 research and innovation programme (project {\sc Spice Dune}, grant agreement No 947634). 
J.L.-B. is funded by the Spanish Ministry of Science and Universities (MICIU/AEI/10.13039/501100011033/) and NextGenerationEU/PRTR grants PID2019-107061GB-C61 and CNS2023-144309. 
This research was funded in part by the UKRI, (Grants ST/X001121/1, EP/X027562/1).
A.C.M.C acknowledges support from the FCT, Portugal, through the CFisUC projects UIDB/04564/2020 and UIDP/04564/2020, with DOI identifiers 10.54499/UIDB/04564/2020 and 10.54499/UIDP/04564/2020, respectively.
This research has made use of the NASA Exoplanet Archive, which is operated by the California Institute of Technology, under contract with the National Aeronautics and Space Administration under the Exoplanet Exploration Program. 
This research has made use of the SIMBAD database \citep{2000A&AS..143....9W}, operated at CDS, Strasbourg, France.
This work has made use of the following software: \texttt{astropy} \citep{2022ApJ...935..167A}, \texttt{matplotlib} \citep{2007CSE.....9...90H}, \texttt{numpy} \citep{2020Natur.585..357H}, and \texttt{scipy} \citep{2020NatMe..17..261V}.

\end{acknowledgements}

%
%

\bibliographystyle{aa} 
\bibliography{references} 
\begin{appendix}

\section{Additional figures}

\begin{figure*}
    \centering
    \includegraphics[width=0.4363\textwidth]{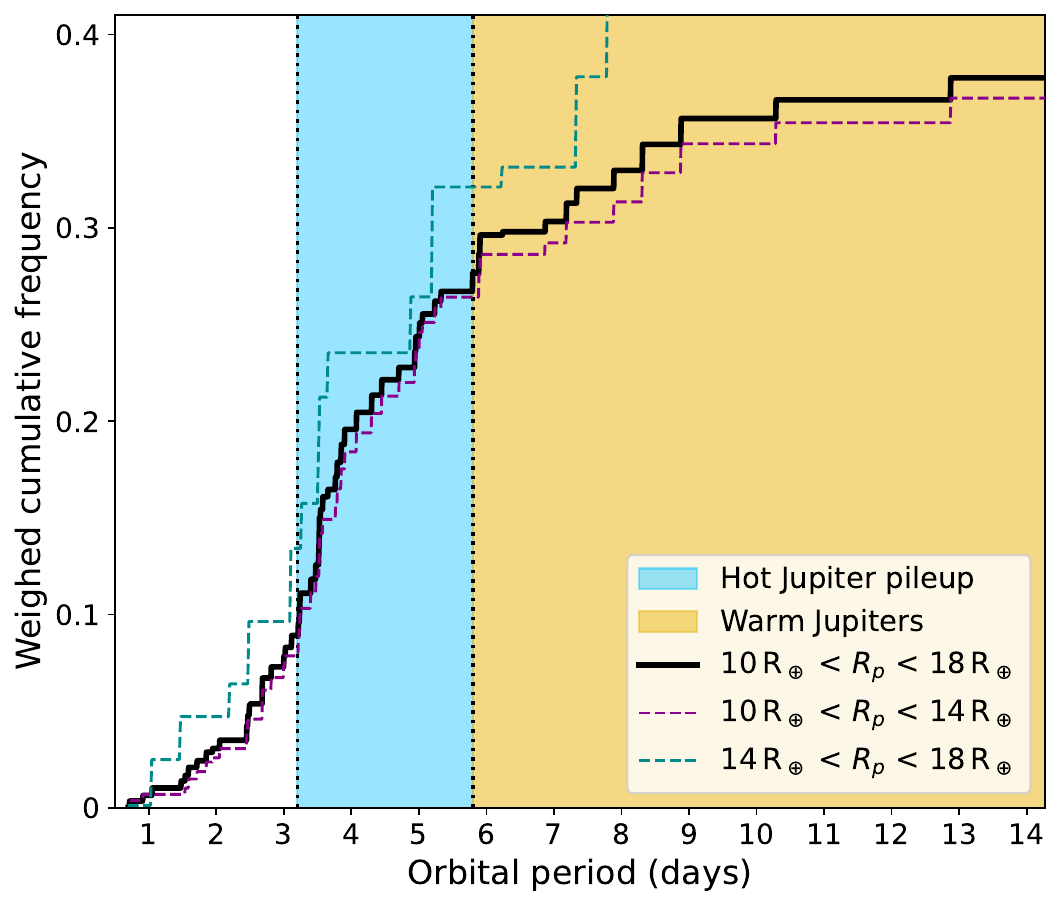}
    \includegraphics[width=0.44\textwidth]{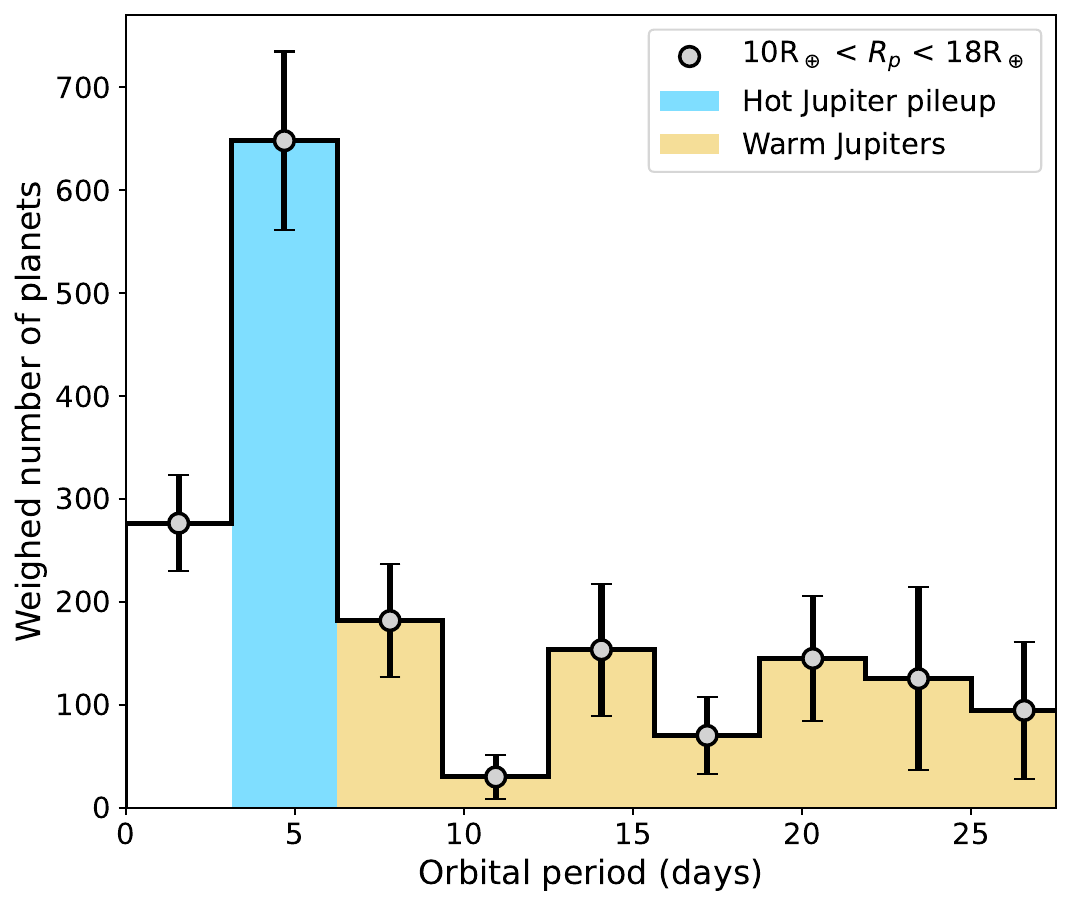}
    \caption{Occurrence of Jupiter-size planets ($R_{\rm p}$ > 10 $\rm R_{\oplus}$) across the orbital period space. The histogram error bars were computed as the square root of the quadratic sum of the weights.}
    \label{fig:oc_jup}
\end{figure*}

\begin{figure*}
    \centering
    \includegraphics[width=0.43\textwidth]{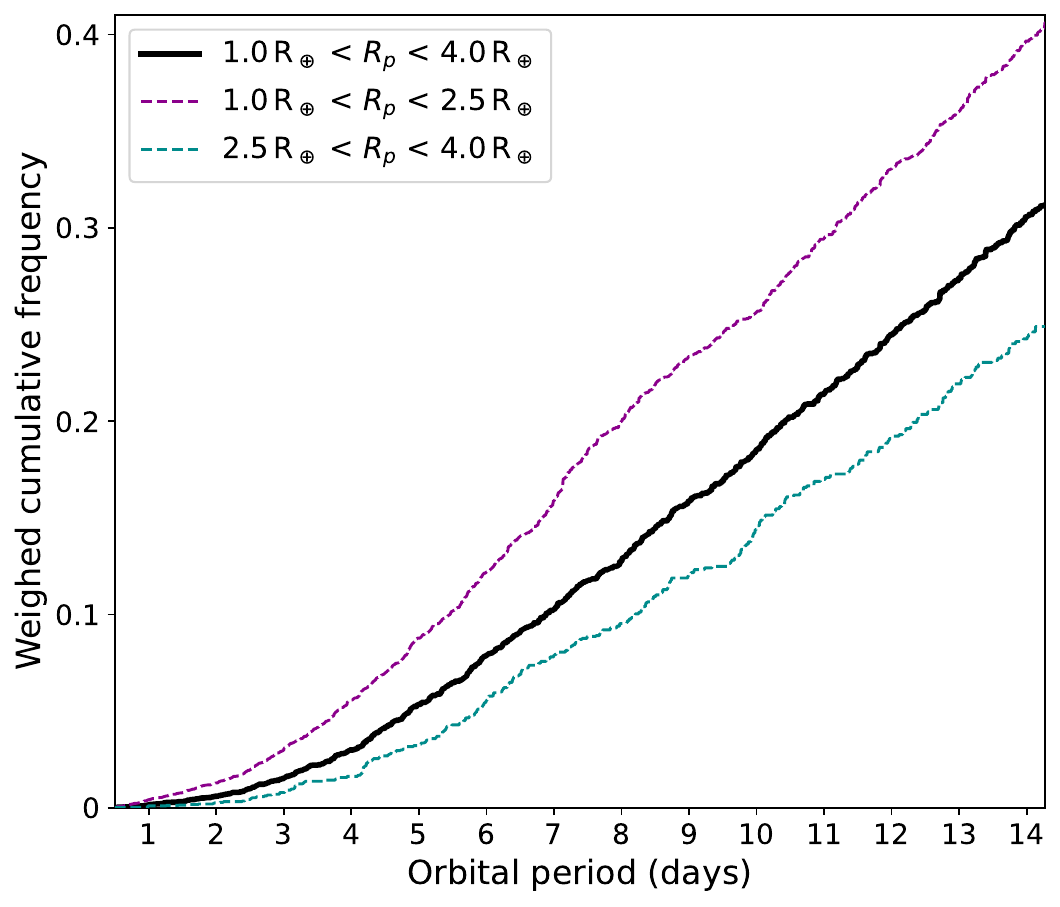}
    \includegraphics[width=0.457\textwidth]{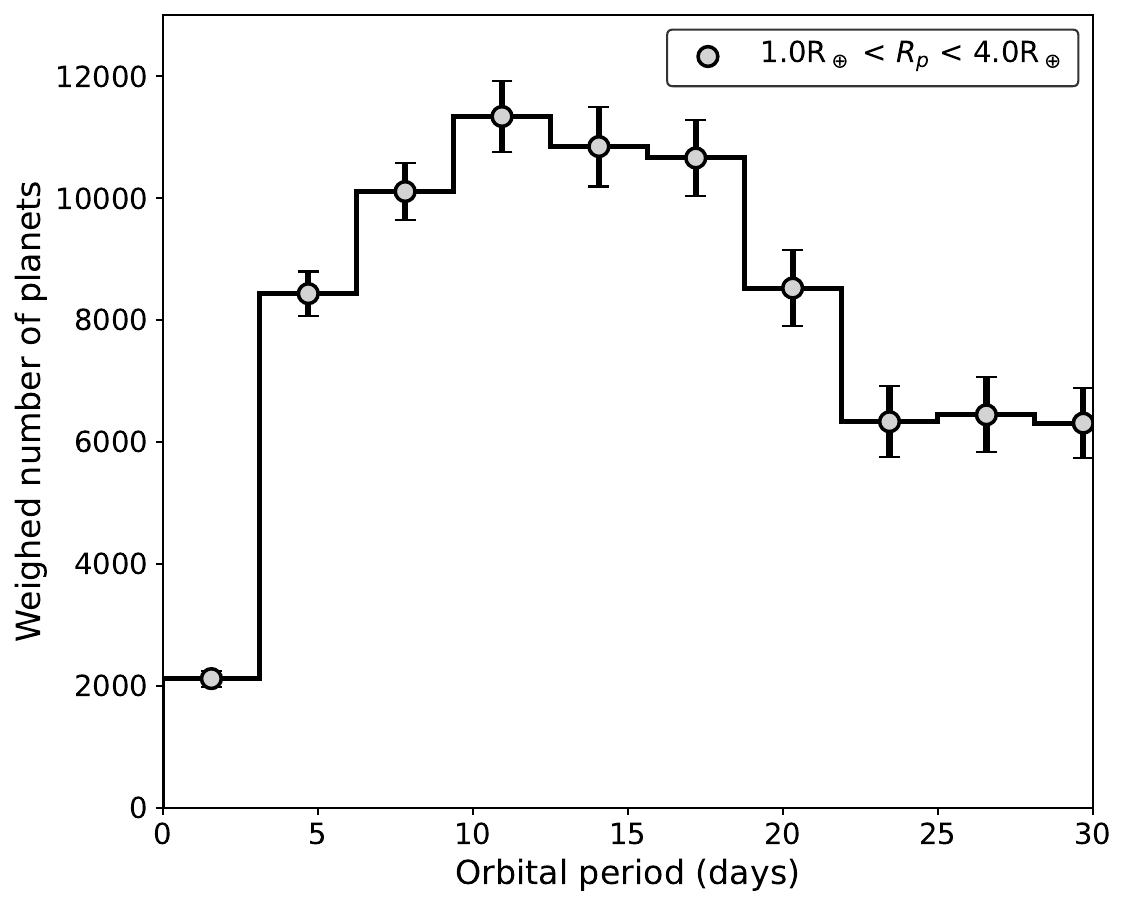}
    \caption{Occurrence of sub-Neptune planets ($R_{\rm p}$ < 4 $\rm R_{\oplus}$) across the orbital period space. The histogram error bars were computed as the square root of the quadratic sum of the weights.}
    \label{fig:oc_sub}
\end{figure*}

\begin{figure*}
    \centering
    \includegraphics[width=0.46\textwidth]{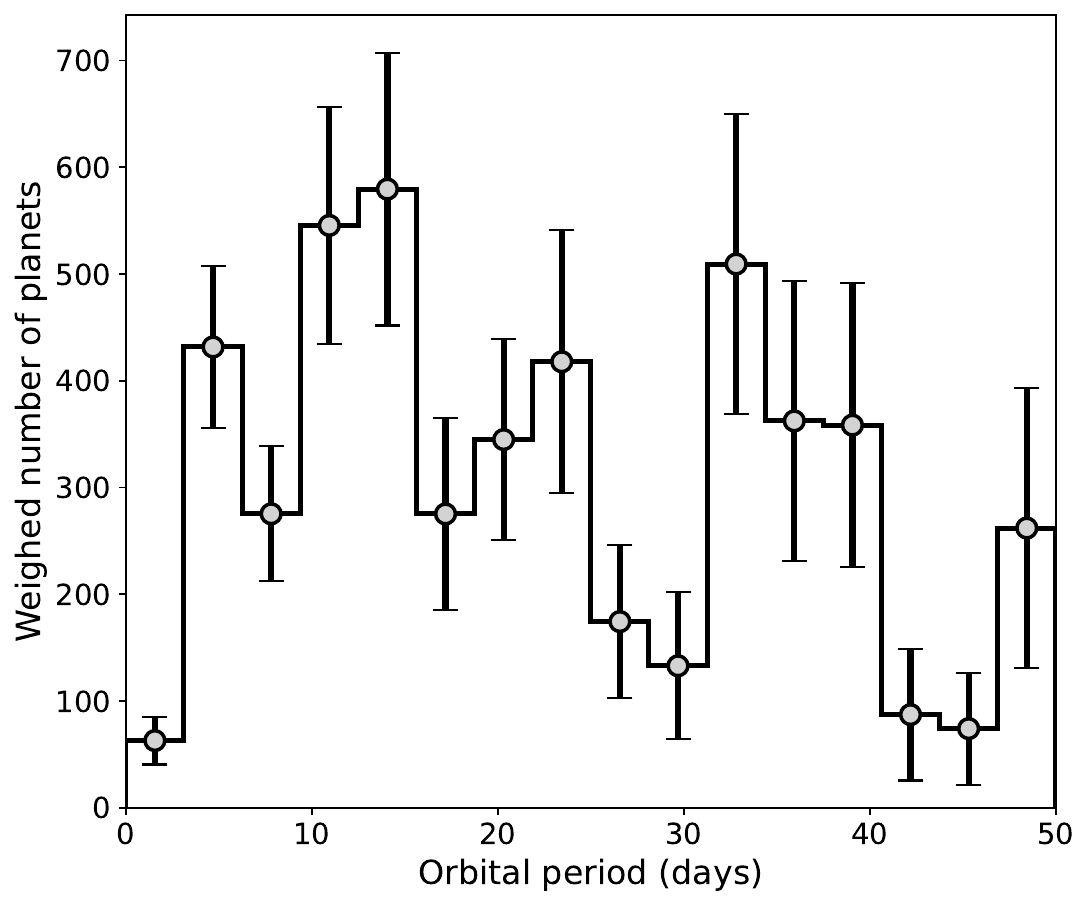}
    \includegraphics[width=0.46\textwidth]{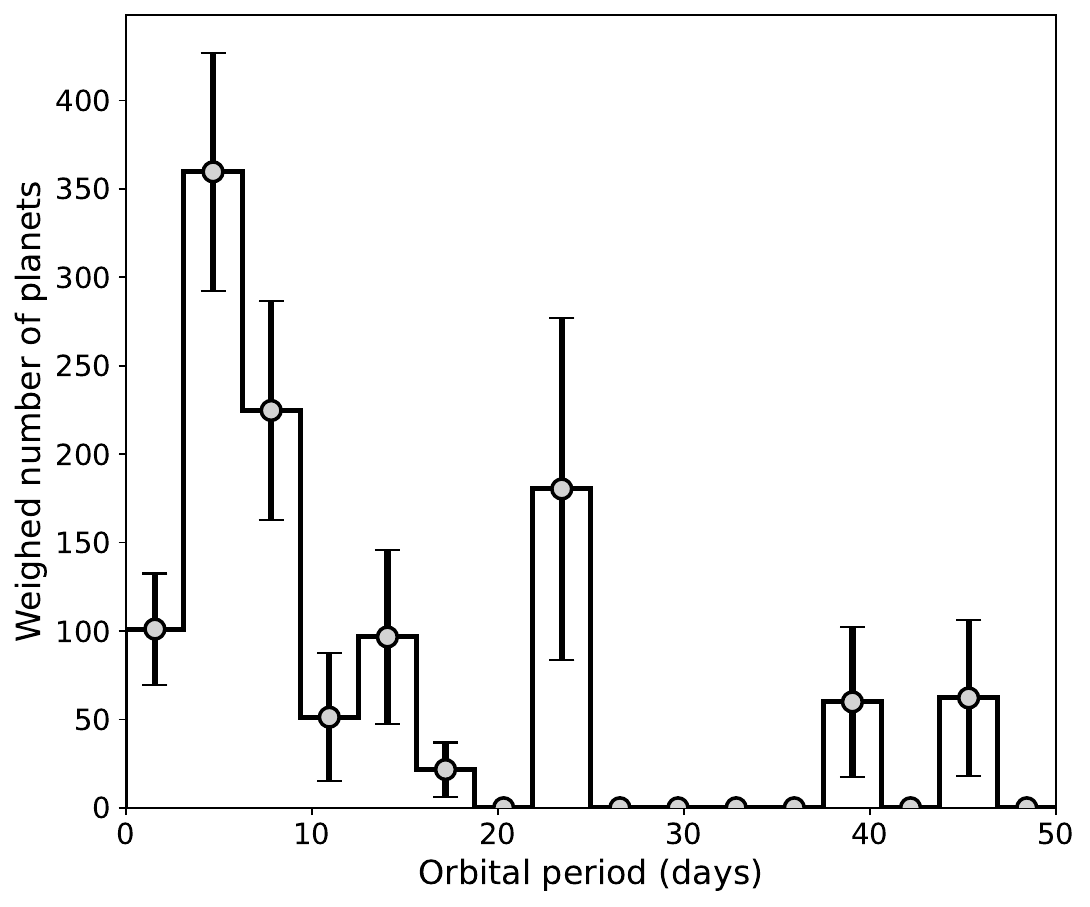}
    \caption{Occurrence of Neptunian planets in the frontier regimes (left panel, 4$\rm R_{\oplus}$ $<R_{\rm p}$ $<$ 5.5$\rm R_{\oplus}$; right panel, 8.5$\rm R_{\oplus}$ $<R_{\rm p}$ $<$ 10$\rm R_{\oplus}$). The histogram error bars were computed as the square root of the quadratic sum of the weights.}
    \label{fig:oc_transitional}
\end{figure*}

\end{appendix}

\end{document}